\documentstyle [12pt] {article}
\title{A SIMPLIFIED, BOHR QUANTUM THEORETICAL DERIVATION OF THE UNRUH TEMPERATURE, ENTROPY AND EVAPORATION}

\author{Vladan Pankovi\'c, Darko V. Kapor\\
Department of Physics, Faculty of Sciences, 21000 Novi Sad,\\ Trg
Dositeja Obradovi\'ca 4, Serbia, \\vladan.pankovic@df.uns.ac.rs}

\date {}
\begin {document}
\maketitle \vspace {0.5cm}
\vspace {0.3cm}

\begin {abstract}
In this work we reproduce Unruh temperature for a spherical
physical system using a simplified rule, very similar to Bohr
angular momentum quantization postulate interpreted via de Broglie
relation, and using a thermal equilibrium stability condition.
(Our rule, by a deeper analysis that goes over basic intentions of
this work, corresponds, for Schwarzschild black hole, to a closed
string loop theory of Copeland and Lahiri.) Firstly, we suppose
that at the surface of a large system gravitational field of this
system can generate some quantum excitations, i.e. small quantum
systems. Mass spectrum of this small quantum system is determined
using a rule that states that circumference of the large system
holds integer numbers of the reduced Compton wavelengths of this
small quantum system. Secondly, we suppose that absolute value of
the classical gravitational interaction between large system and
small quantum system in the ground state is equivalent to thermal
kinetic energy of the small quantum system interacting with large
system as thermal reservoir. It is very similar to virial theorem
of the ideal gas and it yields directly and exactly Unruh
temperature. Finally, using Unruh temperature and thermodynamical
law, we originally propose corresponding Unruh entropy and Unruh
evaporation of the system and demonstrate that for Schwarzschild
black hole Unruh temperature, entropy and evaporation can be
exactly reduced in the Hawking temperature, Bekenstein-Hawking
entropy and Hawking evaporation.
\end {abstract}
\vspace {0.5cm}

As it is well-known, Unruh radiation [1]-[4] represents an
important, general relativistic and quantum effect whose exact
derivation needs a relatively complex theoretical formalism of the
quantum gravity. Some simplifications of the derivation of Unruh
radiation can be done too [5]. In this work "without knowing the
details of quantum gravity" (we paraphrase Fursaev [6]) we shall
reproduce exactly final form of the remarkable Unruh temperature
for a spherical physical system. Firstly, we shall use a
simplified rule, very similar but not identical to Bohr angular
momentum quantization postulate interpreted via de Broglie
relation and by a thermal equilibrium stability condition. (This
rule, by a deeper analysis that goes over basic intentions of this
work, corresponds, for Schwarzschild black hole, to a closed
string loop theory of Copeland and Lahiri [7].) Concretely,
firstly, we suppose that at the surface of a large system
gravitational field of this system (eventually interacting with
quantum vacuum) can generate some quantum excitations, or simply
small quantum systems. Mass (energy) spectrum of this small
quantum system is determined so that circumference of the large
system holds integer numbers of the reduced Compton wavelengths of
this small quantum system. Secondly, we suppose that absolute
value of the classical gravitational interaction between large
system and small quantum system in the ground state is equivalent
to thermal kinetic energy of the small quantum system interacting
with large system as thermal reservoir. It is very similar but not
identical to virial theorem of the ideal gas and it yields
directly and exactly Unruh temperature. Moreover, using Unruh
temperature and thermodynamical law, we originally propose
corresponding Unruh entropy (proportional to product of the system
surface and logarithm of the system mass) and Unruh evaporation of
the system (during time interval approximately inversely
proportional to third degree of the Planck mass. Finally, we
demonstrate that in the especial case when physical system
represents a Schwarzschild black hole Unruh temperature, entropy
and evaporation can be exactly reduced in the Hawking temperature,
Bekenstein-Hawking entropy and Hawking evaporation (time).

Consider a large, massive spherical system with mass M,
characteristic radius R and surface gravity
\begin {equation}
       {\it a}= \frac {GM}{R^{2}}
\end {equation}
where G represents the Newtonian gravitational constant. Suppose
that in the general case there is no functional dependence between
R and M.

Firstly, suppose that at distance R, i.e. at large system surface,
gravitational field of large system (eventually interacting with
quantum vacuum) can create some quantum excitations, or simply -
small quantum systems, whose quantized masses (energies) are
determined by the following condition
\begin {equation}
    m_{n}c R = n \frac {\hbar}{2\pi},
    \hspace{0.6cm} {\rm for} \hspace{0.3cm} m_{n} \ll M \hspace{0.5cm} {\rm and}  \hspace{0.3cm}  n = 1,2,...
\end {equation}
where $m_{n}$ and $ E_{n}= m_{n}c^{2} $for  $ m_{n} \ll M $  and
$ n = 1, 2, … $   represent the mass and energy spectrum of
mentioned small quantum system, while $\hbar$ represents the
reduced Planck constant.

It implies
\begin {equation}
      2\pi R = n \frac {\hbar }{ m_{n}c} = n \lambda_{rn}
      \hspace{0.6cm} {\rm for} \hspace{0.3cm} m_{n} \ll M \hspace{0.5cm} {\rm and}  \hspace{0.3cm}  n = 1,2,...          .
\end {equation}
where $2\pi R$ represents the circumference of the large system,
i.e. circumference of a great circle at large system surface,
while
\begin {equation}
        \lambda_{rn}= \frac {\hbar }{ m_{n}c}
\end {equation}
represents n-th reduced Compton wavelength of the small quantum
system with mass $ m_{n}$ for $n = 1, 2, …$. Expression (3) simply
means that circumference of the large system holds exactly n
corresponding n-th reduced Compton wave lengths of the small
quantum system with mass $ m_{n}$ for $n = 1, 2, …$ .

All this it is very similar to Bohr angular momentum quantization
postulate interpreted via de Broglie relation. But, of course, (2)
cannot be completely consistently interpreted as angular momentum
by simple rotation of the small system along great circle at large
system surface. (Such simple interpretation of (2) as a classical
rotation of the small quantum system leads to rotation speed
$2\pi$ times smaller than c. Namely, if we suppose $m_{1}c= \frac
{\hbar \omega_{1}}{c}$ and $v_{1}= \omega_{1}R$, then, according
to (2), it follows $v_{1}=\frac {c}{2\pi}$.) Simply speaking (2)
represents a postulate on the quantum field theoretical
characteristic of gravitational field of the large system
(eventually interacting with quantum vacuum) at surface of the
system. Or, expression (2) represents a postulate on the effective
reduction of the quantum field theory of gravitational field of
large system at its surface. (Detailed discussion of this
postulate goes over basic intention of this work. We can only
point out that Copeland and Lahiri [7] showed that basic results
of the Schwarzchild black hole thermodynamics can be obtained by
consideration of the small excitations of a closed string loop
which is full agreement with our postulate.)

According to (2) it follows too
\begin {equation}
       m_{n} = n \frac {\hbar}{2\pi Rc}\equiv n m_{1}
       \hspace{0.6cm} {\rm for} \hspace{0.3cm} m_{n} \ll M \hspace{0.5cm} {\rm and}  \hspace{0.3cm}  n = 1,2,...
\end {equation}
\begin {equation}
       m_{1} = \frac {\hbar} {2\pi Rc}
\end {equation}
represents the minimal, i.e. ground mass of the small quantum
system. Obviously, mass spectrum (5), i.e. corresponding energy
spectrum $E_{n}= m_{n}c^{2} = n m_{1}c^{2}= n E_{1}$, for $n = 1,
2, …$ represent practically mass, i.e. energy spectrum of an
linear harmonic oscillator.

Secondly, suppose now that potential energy gravitational
interaction between large system and small quantum system in the
ground state
\begin {equation}
      V= \frac {Gm_{1}M}{R}=\frac {G\hbar M}{2\pi Rc}
\end {equation}
is equivalent to the statistical average kinetic energy of the
small system kT in the contact with large system as thermal
reservoir, i.e. that the following is satisfied
\begin {equation}
      kT = V = \frac {Gm_{1}M}{R}
\end {equation}
where k represents the Boltzmann constant. It, according to (1),
(6), implies
\begin {equation}
      T = \frac {1}{k} \frac {Gm_{1}M}{R} =
      \frac {1}{k} \frac {\hbar}{2\pi c} \frac {GM}{R^{2}} =
      \frac {1}{k} \frac {\hbar}{2\pi c} {\it a}
\end {equation}
representing exactly the Unruh temperature in the general case.

It can be observed that all this is conceptually very similar to
theory of the ideal gas by virial expansion. In this analogy Unruh
temperature corresponds to the temperature of the gas, while small
perturbations of the thermodynamical equilibrium can be described
only by second virial coefficient analogous to Planck law of the
black body radiation or to Bose-Einstein distribution (which
implies that here Stefan-Boltzmann law can be satisfied too).

In the especial case of the Schwarzschild black hole, for which
there is an especial functional dependence between R and M
\begin {equation}
     R = \frac {2GM}{c^{2}}
\end {equation}
representing the Schwarzschild radius, expressions (6), (9), turn
out in
\begin {equation}
       m_{1} = \frac {\hbar c}{4\pi GM}
\end {equation}
\begin {equation}
       T = \frac {\hbar c^{3}}{8\pi GM}      .
\end {equation}
Last expression represents the Hawking temperature.

Further, Unruh radiation must satisfy thermodynamical law
\begin {equation}
      dM c^{2} = T dS
\end {equation}
where $Mc^{2}$ represents total energy of the large system and  S
- corresponding Unruh entropy. It, implies
\begin {equation}
    dS = \frac {1}{T} c^{2} dM
\end {equation}
or, according to (6), (9),
\begin {equation}
   dS = k \frac {2\pi c^{3})}{\hbar} \frac {R^{2}}{GM} dM= \frac{k}{2} \frac {4\pi R^{2}}{\frac {\hbar G}{c^{3}}}  d(\ln [\frac {M}{M_{0}}]) = \frac{k}{2}\frac {A}{L^{2}_{P}} d(\ln[\frac {M}{M_{0}}])
\end {equation}
and
\begin {equation}
   S = \frac{k}{2} \frac {4\pi R^{2}}{\frac {\hbar G}{c^{3}}}  \ln [\frac {M}{M_{0}}] = \frac{k}{2}\frac {A}{L^{2}_{P}} \ln[\frac {M}{M_{0}}]    .
\end {equation}
Here
\begin {equation}
    A = 4\pi R^{2}
\end {equation}
represents the surface of the large system, $L_{P}= (\frac {\hbar
G}{c^{3}})^{\frac {1}{2}}$ - the Planck length and $M_{0}$ - some
constant mass for which it can be supposed that it represents
Planck mass $M_{P}= (\frac {\hbar c}{G})^{\frac {1}{2}}$. In this
way Unruh entropy is pratically proportional to the product of the
large system surface and logarithm of the large system mass. It
represents an original and interesting result.

In the especial case of the Schwarzschild black hole
thermodynamical law, (14), according to (12), implies
\begin {equation}
    dS = \frac {k8\pi GM}{\hbar c}dM = d [{k4\pi GM^{2}}{\hbar c}]
\end {equation}
and
\begin {equation}
    S = {k4\pi GM^{2}}{\hbar c}= \frac {k}{4} \frac {A}{L^{2}_{P}}
\end {equation}
where A represents the Schwarzschild black hole horizon surface
that satisfy (17) for Schwarzschild radius (10). Expression (19)
represents, of course, the Bekenstein-Hawking entropy of the
Schwarzschild black hole.

Suppose that differential form of (14) is changed by corresponding
finite difference form, which according to (6), (9), and (19)
yields
\begin {equation}
    \Delta S_{n} = \frac {1}{T}c^{2}\Delta M = \frac {1}{T} c^{2}m_{n}= n \frac {1}{T} c^{2}m_{1}= n k\frac {Rc^{2}}{GM}
    \hspace{0.6cm} {\rm for} \hspace{0.3cm} m_{n} \ll M \hspace{0.5cm} {\rm and}  \hspace{0.3cm}  n = 1,2,...
\end {equation}
\begin {equation}
    \Delta A_{n} = \frac {4}{k}L^{2}_{P}\Delta S_{n} = n 4 \frac {Rc^{2}}{GM} L^{2}_{P}
    \hspace{0.6cm} {\rm for} \hspace{0.3cm} m_{n} \ll M \hspace{0.5cm} {\rm and}  \hspace{0.3cm}  n = 1,2,... .
\end {equation}
Expressions (20), (21) can be considered as the Unruh entropy and
surface quantization.

In the especial case of the Schwarzschild black hole, according to
(10)-(12), (20), (21) turn out in
\begin {equation}
    \Delta S_{n} =  n 2k
    \hspace{0.6cm} {\rm for} \hspace{0.3cm} m_{n} \ll M \hspace{0.5cm} {\rm and}  \hspace{0.3cm}  n = 1,2,...
\end {equation}
\begin {equation}
    \Delta A_{n}= \frac {4}{k} L^{2}_{P} \Delta S_{n}  = n 8 L^{2}_{P}
    \hspace{0.6cm} {\rm for} \hspace{0.3cm} m_{n} \ll M \hspace{0.5cm} {\rm and}  \hspace{0.3cm}  n = 1,2,... .
\end {equation}
representing Bekenstein quantization  of the black hole entropy
and horizon surface.

Suppose now that large system radiates analogously to the black
body at the Unruh temperature so that Unruh-Stefan-Boltzmann law
is satisfied
\begin {equation}
   -\frac {1}{A}\frac  {dM}{dt} c^{2} = \sigma  T^{4}
\end {equation}
where $\sigma $ represents the Stefan-Boltzmann constant. It
according to (9), (17), implies

\begin {equation}
  -\frac  {dM}{dt}= \frac {\sigma A}{c^{2}} (\frac {1}{k}\frac {\hbar}{2\pi c} \frac {GM}{R^{2}})^{4}=
  \frac {\sigma A}{c^{2}} (\frac {1}{k}\frac {\hbar}{2\pi c}\frac {G}{R^{2}})^{4}M^{4}\equiv \alpha M^{4}
\end {equation}

where $\alpha =\frac {\sigma A}{c^{2}} (\frac {1}{k} \frac
{\hbar}{2\pi c} \frac {G}{R^{2}})^{4}$. After simple integration,
where t increases from initial, zero time moment to final time
moment of the Unruh evaporation $\tau$, while M decrease from
initial mass $M_{in}$ to Planck mass $M_{P}$, (25) yields

\begin {equation}
  \tau = \frac {1}{3\alpha} \frac {M^{3}_{in} - M^{3}_{P}}{ M^{3}_{in}M^{3}_{P}} \simeq \frac {1}{3\alpha M^{3}_{P}}        for M \gg M_{P}      .
\end {equation}
(It can be pointed out that it is chosen that final mass cannot be
smaller than Planck mass, especially that it cannot be zero, since
then Unruh evaporation time tends toward infinity. All this is
full agreement with quantum field theory.) In this way for
practically all macroscopic large system, with mass much larger
than Planck mass, Unruh evaporation time is the same and it is
inversely proportional to the third degree of the Planck mass. It
represents an original and interesting result.

In the especial case of the Schwarzschild black hole, according to
(10), (12), (17), and Hawking-Stefan-Boltzmann law, analogous to
the general form of Unruh-Stefan-Boltzmann law (24) with changed
sign, it follows
\begin {equation}
  \frac {1}{A}\frac  {dM}{dt} c^{2} = \sigma  T^{4}
\end {equation}
which implies
\begin {equation}
  \frac {dM}{dt}= \frac {\sigma}{c^{2}}16\pi  \frac {G^{2}M^{2}}{c^{4}} (\frac {\hbar c^{3}}{k8\pi G})^{4} M^{-4}= \beta M^{-2}
\end {equation}
where $\beta =\frac {\sigma}{c^{2}}16\pi  \frac
{G^{2}M^{2}}{c^{4}} (\frac {\hbar c^{3}}{k8\pi G})^{4}$. After
simple integration, where t increases from initial, zero time
moment to final time moment of the Hawking evaporation $\tau$,
while M decrease from initial mass $M_{in}$ to final mass that
here can be formally used to be zero, (28) yields
\begin {equation}
    \tau = \frac {1}{3\beta}M^{3}_{in}           .
\end {equation}
It means that Hawking evaporation time is proportional to the
third degree of the initial black hole mass. This final result is
principally different from the final result obtained for Unruh
evaporation in the general case (26), but both results are deduced
in the formally completely analogous way.

In conclusion we can shortly repeat and point out the following.
In this work we reproduce exactly final form of the remarkable
Unruh temperature for a spherical physical system using a
simplified rule, very similar but not identical to Bohr angular
momentum quantization postulate interpreted via de Broglie
relation and using a thermal equilibrium stability condition. (Our
rule, by a deeper analysis that goes over basic intentions of this
work, corresponds, for Schwarzschild black hole, to a closed
string loop theory of Copeland and Lahiri.) Concretely, firstly,
we suppose that at the surface of a large system gravitational
field of this system (eventually interacting with quantum vacuum)
can generate some quantum excitations, or simply small quantum
systems. Mass (energy) spectrum of this small quantum system is
determined using a rule that states that circumference of the
large system holds integer numbers of the reduced Compton
wavelengths of this small quantum system. Secondly, we suppose
that absolute value of the classical gravitational interaction
between large system and small quantum system in the ground state
is equivalent to thermal kinetic energy of the small quantum
system interacting with large system as thermal reservoir. It is
very similar but not identical to virial theorem of the ideal gas
and it yields directly and exactly Unruh temperature. Moreover,
using Unruh temperature and thermodynamical law, we originally
propose corresponding Unruh entropy (proportional to product of
the system surface and logarithm of the system mass) and Unruh
evaporation of the system (during time interval approximately
inversely proportional to third degree of the Planck mass.
Finally, we demonstrate that in the especial case when physical
system represents a Schwarzschild black hole Unruh temperature,
entropy and evaporation can be exactly reduced in the Hawking
temperature, Bekenstein-Hawking entropy and Hawking evaporation
(time).

\vspace{0.5cm}

{\large \bf References}

\begin {itemize}

\item [[1]] B. Unruh, Phys. Rev. {\bf D 14} (1976)
\item [[2]] R. M. Wald, {\it Quantum Field Theory in Curved Spacetime and Black Hole Thermodynamics} (University of Chicago press, Chicago, 1994)
\item [[3]] R. M. Wald, {\it The Thermodynamics of Black Holes}, gr-qc/9912119
\item [[4]] D. N. Page, {\it Hawking Radiation and Black Hole Thermodynamics}, hep-th/0409024
\item [[3]]  P. M. Alsing, P. W. Milloni, {\it Simplified Derivation of the Hawking-Unruh temperature for an accelerating observer in vacuum}, Am. J. Phys. 72 (2004) 1524 ; quant-ph/0401170 v2
\item [[6]] D. V. Fursaev, {\it Can One Understand Black Hole Entropy without Knowing Much about Quantum Gravity?}, gr-qc/0404038
\item [[7]] E. J. Copeland, A.Lahiri, Class. Quant. Grav. , {\bf 12} (1995) L113 ; gr-qc/9508031

\end {itemize}

\end {document}